\documentclass{article}
\usepackage[utf8]{inputenc}
\usepackage{xpatch,xcolor}
\usepackage[normalem]{ulem}
\usepackage[margin=1.2in]{geometry}
\usepackage{multicol,multirow}
\usepackage{url,cite}
\usepackage{hyperref}
\hypersetup{
    colorlinks=true,
    linkcolor=purple,
    filecolor=magenta,      
    urlcolor=purple,
}

\title{SASV Challenge 2022: A Spoofing Aware Speaker Verification Challenge Evaluation Plan}
\author{Jee-weon Jung, Hemlata Tak, Hye-jin Shim, Hee-Soo Heo, Bong-Jin Lee,\\ Soo-Whan Chung, Hong-Goo Kang, Ha-Jin Yu, Nicholas Evans, and Tomi Kinnunen}
\date{27 February 2022\footnote{version 0.2.}}

\begin{document}
\maketitle

\section{Introduction}
The performance of automatic speaker verification (ASV) systems has improved dramatically in recent decades~\cite{campbell2006support,dehak2010front,heigold2016end,snyder2018x}.
Even for some \textit{in the wild} scenarios, state-of-the-art systems can deliver equal error rates (EERs) of less than 1\%~\cite{desplanques2020ecapa,thienpondt2021integrating}.
However, these impressive results are typically derived without the consideration of spoofing attacks, namely specially crafted utterances generated by adversaries in order to deceive the ASV system and to provoke false accepts. 
Even state-of-the-art ASV systems can be vulnerable to spoofing attacks generated using speech synthesis / text-to-speech (TTS), voice conversion (VC) or replay attacks.
Some such attacks can degrade ASV reliability considerably~\cite{wang2020asvspoof}.

Led by the ASVspoof initiative and corresponding challenge series, countermeasure (CM) systems have hence been developed in order to help detect and deflect spoofing attacks~\cite{wu2015asvspoof,todisco2019asvspoof,ASV2021challenge}. 
In the case of a logical access, telephony scenario involving only TTS and VC attacks, the best performing spoofing CM systems can deliver EERs of less than 2\%~\cite{lavrentyeva2019stc,chen2020generalization,tak2021end,hua2021towards,li2021channelwise,wang2021comparative,zhang2021effect,luo2021capsule,ge2021raw,jung2021aasist,wang2022practical}. 

This measure of performance only reflects that of the CM, however, whereas it is the reliability of the ASV system which is of primary importance. 
This can remain poor, even when it operates in tandem with a strong CM~\cite{nautsch2021asvspoof}.  
While the minimum tandem detection cost function (t-DCF)~\cite{kinnunen-tDCF-TASLP} reflects the impact of spoofing attacks and CMs upon the ASV system, the ASVspoof challenge series focuses on the development of CMs for a \textbf{fixed} ASV system with a pre-determined operating point. 
We argue that better performance can be delivered when CM and ASV subsystems are both optimised. 
Herein lies the difference between ASVspoof and the new \textbf{S}poofing-\textbf{A}ware \textbf{S}peaker \textbf{V}erification (SASV) challenge. 
SASV extends the focus of ASVspoof upon CMs to the consideration of integrated systems where both CM \emph{and} ASV subsystems are optimised to improve reliability. 

\section{Challenge objectives}
The goal of the new SASV challenge is hence to further improve robustness to both zero-effort impostor access attempts and spoofing attacks by providing a framework to support the optimisation of CM and ASV systems operating in tandem and, ultimately, facilitate the development of single integrated systems.
With only relatively little previous work in this direction~\cite{sizov2015joint,todisco2018integrated,li2020joint,shim2020integrated,gomez2020joint}, the objectives of the first challenge are to:
\begin{itemize}
    \item bridge the gap between the study of ASV and CM systems, and corresponding research communities;
    \item extend the ASV scenario to take spoofing attacks into account;
    \item promote the development of ensemble models towards integrated SASV solutions which operate upon speaker and anti-spoofing embeddings;
    \item encourage the development of single models which have the capacity to reject both utterances spoken by different speakers as well as spoofed utterances. 
\end{itemize}

\section{SASV solutions}
SASV solutions can take the form of two different processing pipelines.

\subsection{Ensemble solutions based upon separate ASV and CM systems}
\label{ssec:ensemble_system}

Ensemble SASV solutions are assumed to comprise pre-trained ASV and CM subsystems.
Different ensemble techniques can be used to combine embeddings/scores produced by the ASV subsystem with embeddings/scores produced by the CM subsystem.

Potential solutions include, e.g.:
\begin{itemize}
    \item score-sum ensembles using cosine similarity scores generated from speaker embeddings produced by a pre-trained ASV subsystem and the scores produced by a pre-trained CM subsystem;
    \item ensemble models which operate upon three different embeddings, namely a pair of speaker embeddings extracted from enrolment and test utterances and a CM embedding.
\end{itemize}

\subsection{Integrated single system solutions}
SASV solutions can also take the form of an integrated, single system. 

Potential solutions include, e.g.:
\begin{itemize}
    \item deep neural networks (DNNs) trained in multi-task fashion using a pair of output layers, namely one for speaker identification and another for spoofing detection;
    \item end-to-end systems with additional objective functions applied to intermediate, hidden layers.
\end{itemize}

\section{Metrics}
SASV performance will be assessed using the classical EER (SASV-EER) as the primary metric.  
Identical to the metrics used in~\cite{sahidullah2016integrated,todisco2018integrated}, the SASV-EER does not distinguish between different speaker (zero-effort, non-target, or impostor) access attempts and spoofed access attempts. 
Additional insights into SASV performance can be gained from comparisons to more traditional estimates of speaker verification performance (SV-EER) estimated from a set of target and non-target trials, in addition to performance when the same system is subjected to spoofing attacks (SPF-EER) whereby non-target trials are replaced with spoofed trials. 
All three EER estimates reflect ASV performance, with both SV-EER and SPF-EER being estimated using different subsets of the full set of trials (i.e., protocol) used for estimating the SASV-EER. 
All SASV metrics are hence different to the EER metric used for ASVspoof challenges. 
The latter is estimated using a CM protocol, not an ASV protocol;
the SPF-EER is measured when an SASV system processes pairs of enrolment and test utterances whereas the EER in the case of ASVspoof challenges is measured when a standalone CM system processes single utterances. 
Table~1 illustrates the ground-truth labels and trial subsets used to measure each of the three different EERs to be used for the SASV challenge. 

\begin{table}[ht]
  \caption{Description of EERs. The system involves enrolment utterance(s) and a test utterance. Enrolment utterance(s) is bona-fide (i.e. genuine) and test utterance belongs to either of the three types.}
  \centering
  \vspace{0.2cm}
  \label{tab:eer_types}
  \begin{tabular}{lccc}
    \hline
    & Target & Non-target & Spoof\\
    \hline
    SV-EER & + & - &  \\
    SPF-EER & + &  & - \\
    SASV-EER & + & - & - \\
    \hline
  \end{tabular}
\end{table}

\section{Protocols}
Participants will be provided with two protocols: 

\begin{itemize}
    \item Development protocol:\\ \texttt{ASVspoof2019\_LA\_asv\_protocol/ASVspoof2019.LA.asv.dev.gi.trl.txt}
    \item Evaluation protocol:\\ \texttt{ASVspoof2019\_LA\_asv\_protocol/ASVspoof2019.LA.asv.eval.gi.trl.txt} 
\end{itemize}

Both protocols, which list target, non-target and spoofed trials, can be downloaded from the SASV 2022 GitHub repository  at~\url{https://github.com/sasv-challenge/SASVC2022_Baseline} or from ASVspoof challenge resources. The first is to be used for the development of SASV solutions. 
The second is to be used only for final performance evaluation.
Both protocols are identical to those used for the ASVspoof 2019 LA challenge, albeit by the organisers for ASV experimentation instead of by participants for CM experimentation. 
This is hence the first time that the two protocols have been used by the participants of any common challenge.
One thing worth noting about the protocols are that multiple enrolment utterances exist for each trial; this would be more familiar to researchers in the ASV community where speaker embeddings from each enrolment utterance is typically averaged to compose the final enrolment speaker embedding. 

\section{Baselines}
We provide two baseline systems, one for each solution strategy described in Section~\ref{ssec:ensemble_system}. 
Each system is based upon the same pre-trained ASV and CM subsystems described further below.
Reproducible software for each subsystem and baseline are also available from the SASV 2022 GitHub repository from which participants can download packages for:
\begin{itemize}
    \item the extraction of speaker embeddings and CM embeddings using corresponding pre-trained subsystems;
    \item the estimation of SASV-EER, SV-EER, and SPF-EER metrics;
    \item Baseline1 and Baseline2 SASV solutions described in Section~\ref{ssec:baseline2}. 
\end{itemize}

\subsection{ASV subsystem}

We adopt the ECAPA-TDNN~\cite{desplanques2020ecapa} pre-trained ASV subsystem using the VoxCeleb2 dataset~\cite{voxceleb2}\footnote{We use the implementation available at \url{https://github.com/TaoRuijie/ECAPATDNN}.}.
The system leverages several recent advances in deep learning, achieves state-of-the-art performance for the VoxCeleb1-O (VoxCeleb1 test set) protocol~\cite{nagrani2017voxceleb} and is widely adopted in the community. 
The architecture is based upon Res2Net with a squeeze-excitation module~\cite{gao2019res2net,hu2018squeeze}.
Participants are referred to~\cite{desplanques2020ecapa} for full details.

Results for the ECAPA-TDNN are illustrated in the first row of Table 2. 
The SV-EER of 0.83\% demonstrates that non-target trials are reliably rejected without causing target trials to be rejected too, even though there is domain mismatch between the ASVspoof data and the VoxCeleb2 data used for ASV training.
The SPF-EER of 29.3\% confirms that a conventional ASV system is vulnerable to spoofing attacks, a result confirmed by the SASV-EER of 22.4\%.

\subsection{CM subsystem}
The baseline CM subsystem is the AASIST model described in~\cite{jung2021aasist}\footnote{We used the implementation available at \url{https://github.com/clovaai/aasist} for CM embedding extraction.}. 
ASVspoof 2019 LA train partition is used for training the system~\cite{wang2020asvspoof}. 
It is based upon an integrated spectro-temporal graph attention network which operates directly on raw-waveform inputs and is used for the extraction of CM embeddings.  
Participants are referred to~\cite{jung2021aasist} for further details.

\begin{table}[ht]
\label{tab:baseline_results}
\caption{
  The three different EERs (\%) for the SASV 2022 development and evaluation partitions.
  SASV-EER for all baselines are calculated using the entire protocol that includes trials used to measure the SV-EER (target vs.\ non-target) and those used to measure the SPF-EER (target vs.\ spoof).
  Results shown for a conventional ASV system (ECAPA-TDNN) and the two baseline solutions.}
\vspace{2mm}
\centerline{
\renewcommand{\arraystretch}{1.4}
\begin{tabular}{lcccccc}
\hline

\multirow{2}{*}{} & \multicolumn{2}{c}{SV-EER}  & \multicolumn{2}{c}{SPF-EER} & \multicolumn{2}{c}{SASV-EER}\\ 
\cline{2-7}
                  & \multicolumn{1}{c}{Dev} & \multicolumn{1}{c}{Eval} &  \multicolumn{1}{c}{Dev} & \multicolumn{1}{c}{Eval}& \multicolumn{1}{c}{Dev} & \multicolumn{1}{c}{Eval}               \\ 
\hline

ECAPA-TDNN & 1.88&1.63 & 20.30&30.75 & 17.38&23.83\\
Baseline1 (score-sum)  &32.88&35.32&0.06&0.67&13.07&19.31\\

Baseline2 (back-end ensemble model) &12.87&11.48&0.13&0.78&4.85&6.37\\
\hline
\end{tabular}}
\end{table}
\subsection{Baseline1: score-sum ensemble}
Baseline1 involves a simple sum of the scores produced by the ASV and CM subsystems. 
Thus, no data is used for this baseline as it does not involve any training nor fine-tuning.
Results are shown in the second row of Table 2.
The Baseline1 SASV-EER of 19.31\% corresponds to a relative reduction of 19\% compared to the ECAPA-TDNN solution (23.83\%).
However, while Baseline1 improves performance when assessment includes spoofed trials, performance in the case of a typical ASV assessment scenario is poor; the SV-EER of the ECAPA-TDNN of 1.63\% increases to 35.32\%. 
This degradation is caused by the summing of non-calibrated scores, each derived using different techniques (e.g., the cosine similarity for ASV scores but the DNN softmax output for CM scores). 
The back-end model ensemble strategy of Baseline2 is proposed as a potentially better solution.

\subsection{Baseline2: back-end model ensemble}
\label{ssec:baseline2}

Baseline2 involves the fusion of three embeddings: one extracted from an ASV enrolment utterance using the ECAPA-TDNN system; a second extracted in identical fashion from a test utterance; a third extracted from the same test utterance using the AASIST spoofing CM. 
The model is a vanilla multi-layer perceptron with three hidden layers, trained using  the ASVspoof 2019 LA train partition. 

\newpage
\section{Training datasets}
\label{sec:train_data}

Participants are permitted to use the following datasets:

\begin{itemize}
    \item ASVspoof 2019 LA train partition~\cite{wang2020asvspoof};
    \item ASVspoof 2019 LA development partition~\cite{wang2020asvspoof};
    \item VoxCeleb~2~\cite{voxceleb2}.
\end{itemize}

While participants can utilise the above data as they wish, it is stressed that ASVspoof 2019 LA train and development partitions were originally intended for the training and development of spoofing CMs. 
Since ground-truth speaker labels are available for the ASVspoof 2019 LA database, it can also be used for the training and development of ASV systems.
The VoxCeleb~2 database was designed for ASV experimentation; it does not contain spoofed data.
The use of data from the ASVspoof 2019 LA evaluation partition for training or development purposes is strictly prohibited.

The ASVspoof 2019 LA dataset can be downloaded at \url{https://datashare.ed.ac.uk/handle/10283/3336}. 
The VoxCeleb2 dataset can be downloaded at \url{https://www.robots.ox.ac.uk/~vgg/data/voxceleb/vox2.html}.
The use of any additional datasets is permitted, but only if they do not contain recording of speech (e.g., Musan~\cite{snyder2015musan} and the Room Impulse Response and Noise Database~\cite{ko2017study}.


\section{Registration process}
Participants are required to register by completing and submitting the registration form available at:
\begin{itemize}
    \item \url{https://forms.gle/htoVnog34kvs3as56}
\end{itemize}

\section{Submission of results}
Each participant/team is requested to submit a system description and their best result (one submission).
Submissions should be made e-mail to \texttt{sasv.challenge@gmail.com} with the subject line ``SASV2022\_challenge\_submission\_\{team name\}''.
A single zip file should be included as an attachment and should contain four files: ``system\_description\_\{team name\}.pdf'', ``results\_\{team name\}.csv'',  ``scores\_dev\_\{team name\}.txt'', and ``scores\_eval\_\{team name\}.txt''

\subsection{System descriptions}
The system description should be formatted according to the INTERSPEECH 2022 paper template.  Submitted system descriptions are expected to report the SASV-EER for both development and evaluation protocols as well as the corresponding SV-EERs and SPF-EERs. 
System descriptions will not be peer-reviewed and there is no page limit.
After the challenge is finished, rankings accompanied by submitted system descriptions will be made publicly available at the challenge website:
\begin{itemize}
    \item \url{https://sasv-challenge.github.io}
\end{itemize}
Participants may choose anonymous team names and may also anonymise their system description.

\subsection{Result}
The ``results\_\{team name\}.csv'' file should include six EERs separated by white space: the SASV-EER for the development set, the SV-EER for the development set, the SPF-EER for the development set, the SASV-EER for the evaluation set, the SV-EER for the evaluation set, and the SPF-EER for the evaluation set.
Score files for both the development and the evaluation protocols named ``scores\_dev\_\{team name\}.txt'' and ``scores\_eval\_\{team name\}.txt'' should contain rows of five entries separated by white space and can be created easily by adding a single column entry containing the classifier score to each row of the protocol file.
Each row contains the following entries: {speaker\_model} {test\_utterance} {attack\_type} {trial\_type} {score}. 

The first three lines of the ``scores\_eval.txt'' file should hence contain entries consistent with the following: \\

\noindent \texttt{LA\_0015 LA\_E\_8147880 bonafide target 0.97}

\noindent \texttt{LA\_0015 LA\_E\_4861467 bonafide target
 0.98}

\noindent \texttt{LA\_0015 LA\_E\_6229989 bonafide target 0.99}\\

\noindent where the last entry in each row should be replaced with participants' system score.

\section{Paper submission}
We aim to present SASV 2022 challenge results at a Special Session at INTERSPEECH 2022\footnote{\url{https://interspeech2022.org}} to which participants are invited to submit their contributions.

\section{Important Dates}
\begin{itemize}
    \item January 19, 2022: Release of evaluation plan
    \item \sout{March 10} \textcolor{red}{March 14}, 2022: Results submission
    \item \sout{March 14} \textcolor{red}{March 18}, 2022: Release of participant ranks
    \item March 21, 2022: INTERSPEECH Paper submission deadline
    \item March 28, 2022: INTERSPEECH Paper update deadline
    \item June 13, 2022: INTERSPEECH Author notification
    \item September 18-22, 2022: SASV challenge special session at INTERSPEECH
\end{itemize}

\bibliographystyle{IEEEbib}
\bibliography{refs.bib}
\end{document}